\documentclass[journal]{IEEEtran}
\usepackage{bm}
\usepackage{graphicx}
\usepackage{caption}
\usepackage{mathrsfs}
\usepackage{amsmath}
\usepackage{array}
\usepackage[table]{xcolor}
\usepackage{color}
\usepackage{epsfig,epstopdf,subfigure}
\usepackage{booktabs}
\usepackage{subfigure}
\usepackage{multirow}

\begin{document}

\title{A Principle Solution for Enroll-Test Mismatch in Speaker Recognition}

\author{Lantian Li, Dong Wang, Jiawen Kang, Renyu Wang, Jing Wu, Zhendong Gao, Xiao Chen
\thanks{
This work was supported by the National Natural Science Foundation of China (NSFC) under the project No.61633013,
and also the Huawei Innovation Research Program under Cross-device Speaker Recognition Project Contract No.YBN2019125091.
Dong Wang is the corresponding author.

L. Li, D. Wang and J. Kang are with the Center for Speech and Language Technologies (CSLT), BNRist
at Tsinghua University, Beijing 100084, China (E-mail: \{lilt,kangjw\}@cslt.org, wangdong99@mails.tsinghua.edu.cn).

R. Wang, X. Chen are with the Noah's Ark Lab, Huawei, at Hong Kong Science Park, Pak Shek kok, Shatin, New Territories, Hong Kong (E-mail: \{wangrenyu1,chen.xiao2\}@huawei.com).

J. Wu, Z. Gao are with Huawei Technologies CO. Limited, Pudong, Shanghai, China (E-mail: wujing14@huawei.com, gaozhendong@hisilicon.com).
}

}

\maketitle

\begin{abstract}

Mismatch between enrollment and test conditions causes serious performance degradation on speaker recognition systems.
This paper presents a statistics decomposition (SD) approach to solve this problem. This approach decomposes the
PLDA score into three components that corresponding to enrollment, prediction and normalization respectively. Given
that correct statistics are used in each component, the resultant score is theoretically optimal.
A comprehensive experimental study was conducted on three datasets with different types of mismatch:
(1) physical channel mismatch, (2) speaking behavior mismatch, (3) near-far recording mismatch.
The results demonstrated that the proposed SD approach is highly effective, and outperforms the
ad-hoc multi-condition training approach that is commonly adopted but not optimal in theory.

\end{abstract}

\begin{IEEEkeywords}
  Speaker Recognition; Deep Speaker Embedding; Condition Mismatch
\end{IEEEkeywords}

\IEEEpeerreviewmaketitle

\section{Introduction}
\label{sec:intro}

Speaker recognition has been emerging as an applicable technique after decades of research~\cite{campbell1997speaker,reynolds2002overview,hansen2015speaker}.
Traditional speaker recognition methods are based on statistical models,
in particular the Gaussian mixture model-universal background model (GMM-UBM)~\cite{Reynolds00}
and the subspace alternatives including joint factor analysis (JFA)~\cite{Kenny07} and the i-vector model~\cite{dehak2011front}.
Recently, deep learning methods have demonstrated significant progress with regard to speaker recognition~\cite{deng2014deep}.
Early research learns frame-level features and derives utterance-level representations by simple average~\cite{ehsan14,li2017deep},
and the following research focuses on deriving utterance-level representations directly~\cite{snyder2018xvector,okabe2018attentive}.
The utterance-level representations are often called \textbf{speaker vectors}, and the process that maps a
variant-length speech signal to a fixed-dimension speaker vector is often called \textbf{speaker embedding}.
As far, the most popular deep learning architecture for speaker embedding is the x-vector model, proposed by Snyder et al.~\cite{snyder2018xvector}.
Compared to an alternative deep learning approach that employs an end-to-end architecture and training scheme~\cite{heigold2016end,zhang2016end,rahman2018attention},
the embedding approach is easier to train~\cite{wang2017need} and the derived speaker vectors can support various speaker-related tasks.
Recent progress on the deep speaker embedding approach includes
more comprehensive architecture~\cite{chung2018voxceleb2,Jung2019raw}, improved pooling methods~\cite{okabe2018attentive,Cai2018,Xie19a,Chen2019tied},
better training criteria~\cite{li2016max,ding2018mtgan,Wang2019centroid,bai2019partial,Gao2019improving,Zhou2019deep}, and
better training schemes~\cite{Li2019boundary,Wang2019phonetic,Stafylakis2019}.


In spite of the tremendous progress, the present speaker recognition performance yet cannot meet the request of real applications.
A particular problem is that the performance of a model trained with data from one domain will be substantially reduced when applied to another domain.
Numerous studies have been carried out to deal with the domain-mismatch problem~\cite{mclaren2011source,shon2017autoencoder,wang2018unsupervised,villalba2012bayesian,
garcia2014supervised,garcia2014unsupervised,glembek2014domain,aronowitz2014inter,aronowitz2014compensating,kanagasundaram2015improving,rahman2018improving}.
Most of the research focuses on domain drift, and the central idea is to build statistical models suitable for the target domain,
by using the knowledge of data or model of the source domain.
These techniques are suitable for solving the mismatch between the training data and the deployment environment.
For example, they can adapt a model trained with a large database of reading speech so as to make it applicable in a new scenario of online meetings.

In this paper, we focus on another mismatch issue that seems more serious: the mismatch between the enrollment and test conditions.
In real applications, the condition of enrollment is often controlled: people tend to speak in the same acoustic environment and in the same manner,
and also the recording device does not change. During test, the condition tends to be much more uncontrollable and unpredictable. It is often significantly different
from the enrollment condition, and varies from one test to another. Significant performance reduction is often observed with this
mismatch~\cite{shriberg2008effects,zhang2018analysis,park2016speaker,afshan2020variable}.
Some typical scenarios that involve serious enrollment-test mismatch are:

\begin{itemize}
\item  Cross-Channel test, where the enrollment uses one device and the test uses another device. This leads to clear mismatch on the recording devices.
\item  Time-Variance test,  where the test is conducted in a few weeks or months after the enrollment. This leads to significant mismatch on the speaking behaviors.
\item  Near-Far field test, where the enrollment uses a near-filed microphone while the test uses a far-field microphone, or vice versa. This leads to clear mismatch on
the recording conditions.
\end{itemize}

More complex situations may involve multiple mismatches.
For example, in the multi-genre scenarios, the speaker may enroll in reading speech with a near electret microphone, while the test
could be in chatting or singing, with a far-field dynamic microphone.

The essential problem associated with this enrollment-test mismatch is that the statistical properties
(e.g., between-speaker variance and within-speaker variance) of the enrollment data and the test data are different,
which we call \textbf{statistics incoherence}.
This incoherence fails most of the
scoring models, e.g., the famous probabilistic linear discriminant analysis (PLDA)~\cite{Ioffe06,prince2007probabilistic},
for which statistics coherence between enrollment and test is a prerequisite.


Inter-dataset variability compensation (IDVC)~\cite{aronowitz2014inter} and joint PLDA~\cite{ferrer2019joint,ferrer2018generalization}
model speaker and condition as two variables, and compensate for the variance associated with condition during the inference. In particular,
joint PLDA can compensate for enrollment-test mismatch by specifying that the conditions of the enrollment and test are different.
These \emph{factorization} models rely on multi-conditional data to separate the variations associated with speaker and condition, which is
not possible for a particular enrollment-test condition pair where the condition factor could be vague. Moreover, the compensation
does not fully utilize the statistical information in each condition, so is only partial.

Another commonly used approach in practice is score calibration~\cite{brummer2006application,mclaren2014trial,mandasari2013quality}. This approach
normalizes the score distribution according to the test condition, making the decision robust to condition change.
For a particular enrollment-test condition pair, however, calibration does not improve the discriminant capacity of the score.


This paper will present a novel solution for the enrollment-test mismatch problem,
especially for \emph{a particular enrollment-test condition pair}.
{
The solution decomposes the PLDA score into three components:
enrollment, prediction, and normalization. By this composition, different statistical models
can be used for different components. If we use the statistics of the enrollment condition
for the enrollment component, and the statistics of the test condition for the prediction and
normalization components, the resultant PLDA score remains sound even if the enrollment and test
conditions are mismatched.
}
We call this approach \textbf{statistics decomposition (SD)}. It is a principle solution
for the enrollment-test mismatch problem, as the resultant score is optimal if the statistics on
the enrollment and test conditions are both accurate, even if they are not coherent.

The rest of the paper is organized as follows. Section~\ref{sec:theory} will present a theoretical background by revisiting the
PLDA model, especially the decomposition form.
Section~\ref{sec:method} presents the SD approach for the enrollment-test mismatch problem based on a simple linear transformation.
Experiments are reported in Section~\ref{sec:exp}, and finally the paper is concluded in Section~\ref{sec:con}.

\section{Theory background}
\label{sec:theory}

\subsection{Revisit PLDA}

Giving some speaker vectors $\pmb{x}_1, \pmb{x}_2, .., \pmb{x}_n$ for enrollment and a speaker vector $\pmb{x}$ for test,
the task of speaker verification is to check which of the following two hypotheses is more probable:

\begin{itemize}
\item $H_0$: $\pmb{x}$ and $\pmb{x}_1, \pmb{x}_2, .., \pmb{x}_n$  are produced from the same speaker class;
\item $H_1$: $\pmb{x}$ and $\pmb{x}_1, \pmb{x}_2, .., \pmb{x}_n$  are produced from different speaker classes.
\end{itemize}

Following the Bayes theory, decisions based on the following log likelihood ratio (LLR) are optimal, in terms of
minimum Bayes risk (MBR):

\begin{equation}
\text{LLR} = \log \frac{p(\pmb{x}, \pmb{x}_1, ..., \pmb{x}_n|H_0)}{p(\pmb{x}, \pmb{x}_1, ..., \pmb{x}_n|H_1)}.
\end{equation}

Applying the assumptions of $H_0$ and $H_1$, the LLR score can be written as follows:

\begin{equation}
\label{eq:llr}
\text{LLR} = \log \frac{p(\pmb{x},\pmb{x}_1,...,\pmb{x}_n)}{p(\pmb{x}) p(\pmb{x}_1,...,\pmb{x}_n)},
\end{equation}
\noindent

\noindent where $p(\pmb{x}_1, \pmb{x}_2, ..., \pmb{x}_n)$ is the joint probability of
$\pmb{x}_1, \pmb{x}_2, ..., \pmb{x}_n$ under the assumption that they are from the same speaker.
Note that the numerator can not be factorized further as the variables are from the same speaker and so are dependent on each other.

PLDA assumes a linear Gaussian model for the generation process of $\pmb{x}$, where both the between-class and
within-class distributions are Gaussians.
In this work, we assume that both the between-speaker and within-speaker covariances are full-rank,
which is known as a \emph{two-covariance model}~\cite{brummer2010speaker}. More discussion about this
special form and other PLDA variants can be found in~\cite{sizov2014unifying}.

To simplify the presentation, we further assume that a linear transformation has been
applied to the data so both the between-speaker and within-speaker covariances are diagonal.
Moreover, we will use the isotropic form for both the two covariances in the following formulation.
In practice, the usual case is that at least one of the covariances is non-isotropic, but the
formulation is very similar to the all-isotropic form presented below.

Let $\pmb{\mu}$ denote the class mean, the PLDA model, in the form of two isotropic covariances,
can be written as follows:

\begin{equation}
\label{eq:pmu}
p(\pmb{\mu}) = N(\pmb{\mu}; \pmb{0}, \epsilon \mathbf{I});
\end{equation}

\begin{equation}
\label{eq:px-mu}
p(\pmb{x}|\pmb{\mu}) = N(\pmb{x}; \pmb{\mu}, \sigma \mathbf{I}),
\end{equation}

\noindent where $\epsilon$ and $\sigma$ are the between-speaker variance and the within-speaker variance, respectively.

With this model, it is easy to derive the marginal probability $p(\pmb{x})$ as follows:

\begin{equation}
\label{eq:px}
p(\pmb{x}) = N(\pmb{x}; \pmb{0}, (\epsilon + \sigma) \mathbf{I}),
\end{equation}

\noindent and the marginal probability for multiple vectors of a single speaker can be obtained as follows:

\begin{eqnarray}
p(\pmb{x}_1,...,\pmb{x}_n) &\propto  \exp\big \{ - \frac{1}{2\sigma} \{ \sum_i  ||\pmb{x}_i||^2 - \frac{n^2 \epsilon }{\sigma+n\epsilon}||\bar{\pmb{x}}||^2 \} \big\}.
\end{eqnarray}

With this marginal, it is easy to compute LLR according to Eq.(\ref{eq:llr}).

\subsection{Decomposition form of PLDA}

The LLR score can be formulated into a decomposition form~\cite{kenny2013plda,borgstrom2013discriminatively,mccree2017extended,wang2020remark,wang2020simulation} as follows:

\begin{eqnarray}
\text{LLR} &=& \log \frac{p(\pmb{x}|\pmb{x}_1,...,\pmb{x}_n)}{p(\pmb{x})} \nonumber \\
    &=& \log \int p(\pmb{\mu} | \pmb{x}_1, ..., \pmb{x}_n) p(\pmb{x} | \pmb{\mu})\, \rm{d}{\pmb{\mu}} - \log \mathit{p}(\pmb{x}). \label{eq:llr-dec}
\end{eqnarray}

By this formulation, the LLR scoring can be separated to three phases:

\begin{itemize}
\item \textbf{Enrollment}: Compute the posterior $p(\pmb{\mu}|\pmb{x}_1, \pmb{x}_2, ..., \pmb{x}_n)$.
\item \textbf{Prediction}: Compute the likelihood $p_k(\pmb{x})=\int p(\pmb{\mu} | \pmb{x}_1, \pmb{x}_2, ..., \pmb{x}_n) p(\pmb{x} | \pmb{\mu}) \rm{d}{\pmb{\mu}}$.
\item \textbf{Normalization}: Compute the normalization term $p(\pmb{x})$ and use it to normalize the likelihood by $\log p_k(\pmb{x}) - \log p(\pmb{x})$.
\end{itemize}

This decomposition offers an intuitive interpretation for the LLR score, and simplifies the computation. According to the PLDA model, the posterior for $\pmb{\mu}$ is
computed as follows:

\begin{equation}
\label{eq:pmu-xx}
 p(\pmb{\mu}|\pmb{x}_1,...,\pmb{x}_n) = N(\pmb{\mu}; \frac{n\epsilon}{n\epsilon + \sigma} \bar{\pmb{x}}, \frac{\epsilon \sigma}{n\epsilon + \sigma} \mathbf{I}),
\end{equation}

\noindent where $\bar{\pmb{x}}$ is the average of the observations $\pmb{x}_1,...,\pmb{x}_n$.
The likelihood $p(\pmb{x}|\pmb{x}_1,...,\pmb{x}_n)$ can therefore be computed by marginalizing over $\pmb{\mu}$:

\begin{eqnarray}
p(\pmb{x}|\pmb{x}_1,...,\pmb{x}_n) &=& \int p(\pmb{x}| \pmb{\mu}) p(\pmb{\mu}|\pmb{x}_1,...,\pmb{x}_{n}) \rm{d} \pmb{\mu} \nonumber \\
             &=& N(\pmb{x}; \frac{n\epsilon}{n\epsilon + \sigma} \bar{\pmb{x}}, (\sigma + \frac{\epsilon \sigma}{n\epsilon + \sigma})\mathbf{I}). \label{eq:test}
\end{eqnarray}

A simple computation shows that:

\begin{eqnarray}
\text{LLR} &=& \log p(\pmb{x}|\pmb{x}_1,...,\pmb{x}_n) - \log p(\pmb{x}) \label{eq:nl-unknown-log} \\
                           &\propto& -\frac{1}{\sigma + \frac{\epsilon \sigma}{n \epsilon + \sigma}} ||\pmb{x} - \tilde{\pmb{\mu}}||^2 + \frac{1}{\epsilon + \sigma} ||\pmb{x}||^2, \nonumber
\end{eqnarray}
\noindent where we have defined:

\begin{equation}
\tilde{\pmb{\mu}} = \frac{n\epsilon}{n\epsilon + \sigma} \bar{\pmb{x}}.
\end{equation}

\section{Statistics Decomposition}
\label{sec:method}

In this section, we employ the decomposition form of PLDA to address enrollment-test mismatch. Note that the essential problem in the enrollment-test mismatch scenario
is that the statistics of the enrollment and test conditions are incoherent, and so require different models to represent each of them.
The standard form of PLDA LLR involves the enrollment and test data in a single model,
as evidenced by the joint likelihood $p(\pmb{x}, \pmb{x}_1, ..., \pmb{x}_n)$. This makes it unsuitable for dealing with enrollment-test mismatch.
In contrast, the decomposition form decouples the scoring process into separate phases, and each phase can
use its own (and correct) statistical model. This offers a principle solution for the enrollment-test mismatch problem, which we call \textbf{statistics decomposition (SD)}.

\subsection{General form}
\label{sec:gf}

With the SD approach, it is clear that the enrollment and normalization phases should be based on the statistics of the enrollment and
test condition, respectively. The prediction phase, however, it is more complex. According to Eq.(\ref{eq:llr-dec}),
it marginalizes the conditional probability $p(\pmb{x}|\pmb{\mu})$ with respect to the posterior $p(\pmb{\mu}|\pmb{x}_1, ..., \pmb{x}_n)$.
However, $p(\pmb{\mu}|\pmb{x}_1, ..., \pmb{x}_n)$ is computed based on the statistical model of the enrollment condition,
and so is not a correct posterior for marginalizing the conditional probability $p(\pmb{x}|\pmb{\mu})$ where $\pmb{\mu}$ and $\pmb{x}$
are quantities in the test condition.

In order to solve this problem, a connection must be established between the quantities in the enrollment condition and the test condition.
A possible solution is to design a mapping function that transforms $\pmb{\mu}$ from the enrollment condition
to the test condition, or vice versa, transforms $\pmb{x}$ from the test condition to the enrollment condition.
In this study, we will choose the transformation on $\pmb{x}$, as it results in a relatively simpler form.
For simplicity, we shall assume that the transformation is linear:

\begin{equation}
\label{eq:linear}
\pmb{x} = \mathbf{M} \hat{\pmb{x}} + \pmb{b},
\end{equation}
\noindent where $\hat{\pmb{x}}$ represents the observation in the test condition, and $\pmb{x}$ is the transformed data in the enrollment condition.
If we assume that the transformed data can be well represented by the statistical model of the enrollment condition,
the LLR score can be derived.

Firstly, denote the enrollment data of the $k$-th class by $\pmb{x}^k_1,...,\pmb{x}^k_{n_k}$.
According to Eq.~(\ref{eq:pmu-xx}), the posterior $p(\pmb{\mu}_k |\pmb{x}^k_1,...,\pmb{x}^k_{n_k} )$ is computed using the model of the enrollment condition:

\begin{equation}
p(\pmb{\mu}_k |\pmb{x}^k_1,...,\pmb{x}^k_{n} ) = N(\pmb{\mu}_k ;  \frac{n_k\epsilon}{n_k\epsilon + \sigma} \bar{\pmb{x}}_k, \frac{\epsilon \sigma}{n_k\epsilon + \sigma} \mathbf{I}).
\end{equation}

Secondly, transform the test sample $\hat{\pmb{x}}$  by the linear transformation, and perform the prediction using the model of the enrollment condition:

\begin{eqnarray}
\label{eq:pkx-test}
p_k(\hat{\pmb{x}}; \mathbf{M}, \pmb{b}) &=&p(\mathbf{M} \hat{\pmb{x}} + \pmb{b}|\pmb{x}^k_1,...,\pmb{x}^k_{n}) \nonumber \\
             &=& \int p(\mathbf{M} \hat{\pmb{x}} + \pmb{b}|\pmb{\mu}_k) p(\pmb{\mu}_k|\pmb{x}^k_1,...,\pmb{x}^k_{n}) \rm{d} \pmb{\mu}_k \nonumber \\
             &=& N(\mathbf{M} \hat{\pmb{x}} + \pmb{b}; \frac{n_k\epsilon}{n_k\epsilon + \sigma} \bar{\pmb{x}}_k, (\sigma + \frac{\epsilon \sigma}{n_k\epsilon + \sigma})\mathbf{I}). \nonumber \\
\end{eqnarray}

According to Eq.~(\ref{eq:px}), the normalization term $p(\hat{\pmb{x}})$ is computed based the model of the test condition:

\begin{equation}
\label{eq:px-test}
p(\hat{\pmb{x}}) = N(\hat{\pmb{x}}; \pmb{0}, (\hat{\epsilon} + \hat{\sigma}) \mathbf{I}).
\end{equation}

The LLR score is then computed as follows, by using the results of Eq.~(\ref{eq:pkx-test}) and Eq.~(\ref{eq:px-test}):

\begin{equation}
\label{eq:nl-gf}
\text{LLR} (\hat{\pmb{x}}|k) \propto -\frac{1}{\sigma + \frac{\epsilon \sigma}{n_k \epsilon + \sigma}} ||\mathbf{M} \hat{\pmb{x}} + \pmb{b} - \tilde{\pmb{\mu}}_k||^2 + \frac{1}{\hat{\epsilon} + \hat{\sigma}} ||\hat{\pmb{x}}||^2,
\end{equation}

\noindent where we have defined:
\begin{equation}
\label{eq:muk}
\tilde{\pmb{\mu}}_k = \frac{n_k\epsilon}{n_k\epsilon + \sigma} {\pmb{x}}_k.
\end{equation}

The optimal parameters $\{\mathbf{M}, \pmb{b}\}$ for the linear transformation
can be estimated by maximum likelihood (ML) training~\cite{myung2003tutorial},
which maximizes $p_k(\hat{\pmb{x}}; \mathbf{M},\mathbf{b})$ with respect to $\mathbf{M}$ and $\mathbf{b}$,
using samples in the test condition and the samples from the same speaker in the enrollment condition.
It is essentially a maximum likelihood linear regression (MLLR) task~\cite{leggetter1995maximum}, and the
objective function for the optimization can be written by:

\begin{equation}
\label{eq:mle-loss}
\mathcal{L}(\mathbf{M}, \pmb{b}) = \sum_{k=1}^{K} \sum_{i=1}^{N} p_k(\hat{\pmb{x}}_i; \mathbf{M}, \pmb{b}),
\end{equation}
\noindent where $K$ denotes the number of speakers, and $N$ denotes the number of test samples in each speaker.

Note that with the ML training, we essentially minimize the KL divergence between the within-speaker distribution
of the enrollment condition and the one of the test condition after the linear transformation.
If the two within-speaker distributions are perfectly matched with the linear transformation in Eq.(\ref{eq:linear}),
the LLR score computed as above will be MBR optimal.
This approach will be called \textbf{statistics decomposition with linear transformation (SD/LT)} in the paper.

We highlight that if the condition mismatch is very complex, a more complicated transformation is required.
In any case, if the within-speaker distributions in the two conditions can be perfectly matched with the transformation,
the LLR score will be MBR optimal. We therefore obtain a principle solution for the enrollment-test mismatch problem.
As a preliminary study, we focus on SD/LT approach in this paper and leave more complicated transformations (e.g., those based on
deep neural nets) as future work.

\subsection{Special case 1: global shift compensation}
\label{sec:gsc}

A special case of the enrollment-test mismatch is that the behavior of the speakers do not change much
(i.e. assume the between- and within-speaker variances do not change between the enrollment and test conditions),
but the speaker vectors are globally shifted by a constant vector $\pmb{b}$.
This often happens in channel mismatch, where a displacement is often observed between the speaker vectors recorded from
devices with different physical characteristics.

In this special case, we assume the between-speaker and within-speaker covariances do not change,
therefore a simple global shift $\pmb{b}$ on the speaker vectors of the test condition is sufficient to
obtain the optimal LLR score. Formally, the SD/LT is simplified to be:

\begin{equation}
\pmb{x} = \hat{\pmb{x}} + \pmb{b},
\end{equation}

\noindent and the LLR score is derived by:

\begin{equation}
\label{eq:nl-gsc}
\text{LLR} (\hat{\pmb{x}} |k ) \propto  -\frac{1}{\sigma + \frac{\epsilon \sigma}{n_k \epsilon + \sigma}} ||\hat{\pmb{x}} + \pmb{b} - \tilde{\pmb{\mu}}_k ||^2 + \frac{1}{\epsilon + \sigma} ||\hat{\pmb{x}} + \pmb{b}||^2,
\end{equation}

\noindent where $\tilde{\pmb{\mu}}_k$ is defined by Eq.(\ref{eq:muk}).

\subsection{Special case 2: within-speaker variance adaptation}
\label{sec:wva}

As the second special case, considering that people tend to speak with a larger variation during test than during enrollment.
This often happens when a registered person performs time-variance test (e.g., several days or weeks later) and the
test is performed multiple times. In this case, the speaking behavior at each test tends to be different, thus leading to a large
within-speaker variation.

For this special case, we will assume that the within-speaker variance is the only changed statistical
quantity\footnote{One may argue that the speaking behavior at each test time tends to be the same, and so the
variation of each time is not large. This can therefore be treated as a global shift as in channel mismatch.
However, this is not applicable because the speaking behavior at each test time cannot be determined, and what we can
obtain is only the long-term change on the within-speaker variation. }.
With this assumption, it is easy to compute the LLR score for a test sample $\pmb{x}$ by using $\sigma$
in the enrollment phase, and $\hat{\sigma}$ in the prediction phase and normalization phase:

\begin{equation}
\label{eq:nl-wva}
\text{LLR} (\pmb{x} |k ) \propto  -\frac{1}{\hat{\sigma} + \frac{\epsilon \sigma}{n_k \epsilon + \sigma}} ||\pmb{x} -  \tilde{\pmb{\mu}}_k ||^2 + \frac{1}{\epsilon + \hat{\sigma}} ||\pmb{x}||^2,
\end{equation}

\noindent where $\tilde{\pmb{\mu}}_k$ is defined by Eq.(\ref{eq:muk}).

\section{Related work}
\label{sec:rel}

A multitude of methods have been developed for addressing \emph{training-deployment domain mismatch}.
These algorithms can be categorized to \emph{data theme}~\cite{wang2018unsupervised,shon2017autoencoder},
\emph{model theme}~\cite{mclaren2011source,glembek2014domain,Williams2019ff,bhattacharya2019adapting,kang2020disentangled,zhang2018analysis}
and \emph{scoring theme}~\cite{aronowitz2014inter,aronowitz2014compensating,kanagasundaram2015improving,rahman2018improving,villalba2012bayesian,garcia2014supervised,garcia2014unsupervised,lee2019coral},
working on data/features, embedding models and scoring models respectively.
For each theme, the basic idea is either normalization or adaptation: the former aims to make the data or model
domain-independent~\cite{mclaren2011source,wang2018unsupervised,Williams2019ff,bhattacharya2019adapting,kang2020disentangled,aronowitz2014inter,aronowitz2014compensating,kanagasundaram2015improving,rahman2018improving}, while the latter aims to make them more suitable for the target domain~\cite{shon2017autoencoder,glembek2014domain,zhang2018analysis,villalba2012bayesian,garcia2014supervised,garcia2014unsupervised,lee2019coral}.
The normalization methods can be readily applied to alleviate the \emph{enrollment-test mismatch problem},
though the adaptation methods can not as the enrollment and test are in different conditions.

Multi-condition training (MCT)~\cite{afshan2020variable,lei2012towards,rajan2013effect} is among the mostly used approach
to enrollment-test mismatch. It can be regarded as a special normalization method,
belonging to the scoring theme. It pools the data from both enrollment and test conditions and trains a multi-conditional PLDA.
Intuitively, it emphasizes on the most speaker-relevant variation and attenuates other speaker-irrelevant variations,
and so can alleviate impact from condition mismatch.
This approach works well in many scenarios, however it does not offer an optimal score,
as its statistical model is neither accurate for the enrollment condition nor for the test condition.
Moreover, for a particular enrollment-test pair, the condition factor might be unknown and uncovered by the training data.


IDVC~\cite{aronowitz2014inter} is another possible approach, aiming at factorizing speaker variance and condition variance
in the speaker vector space.
Joint PLDA~\cite{ferrer2019joint,ferrer2018generalization} holds a similar idea, though formulates it as a more complicated statistical model.
By this model, speaker and condition variations are represented by two Gaussian distributions, and their inter-correlation
is represented by a linear combination. This probabilistic formulation offers a theoretically sound compensation for
enrollment-test mismatch.
However, the compensation is \emph{partial}, i.e., it only compensates for `the fact that the mismatch exists', but not for the mismatch itself,
i.e., the distributional discrepancy between enrollment and test.
Moreover, training the model requires data to cover different values of well-defined condition factors (e.g., 16 languages in~\cite{ferrer2019joint}).
This is not suitable for our purpose of addressing \emph{any} enrollment-test pair where the condition factor could be vague.

Various score normalization techniques have been demonstrated to be effective to tackle complex conditions,
for example the adaptive symmetric norm (AS-norm)~\cite{matejka2017analysis}.
In the enrollment-test mismatch scenario, AS-norm can partly alleviate the bias caused by the mismatched statistical model. This is
achieved by employing statistics of imposter scores that posses the same bias. AS-norm and SD/LT work in two different
domains (AS-norm on scores and SD/LT on models) and are orthogonal and complementary. We will report potential improvement when
AS-norm is applied together with SD/LT.

Finally, score calibration~\cite{brummer2006application,mclaren2014trial,mandasari2013quality} is another score-domain approach that has
been popular used in practice. This approach basically
adjusts the global distribution of the scores from the recognition system (e.g., PLDA output) to ensure that the calibrated score
represents LLR. This can improve the global performance of practical systems in multi-conditional scenarios.
However, for a particular enrollment-test pair where the test data tend to be homogeneous (in terms of
the calibration function), calibration generally does not improve the discriminative capacity of the score, and therefore does not impact
the EER results, although it certainly boosts performance in calibration-related metrics such as $C_{llr}$~\cite{brummer2006application}.
Note that if the test data is indeed heterogeneous, trial-based calibration~\cite{mclaren2014trial} may improve the EER performance. In this case,
calibration is an orthogonal technique as the proposed SD/LT approach.

\section{Experiments}
\label{sec:exp}

We will conduct experiments on three datasets to verify the proposed SD/LT scoring approach. The three tests
reflect different types of enrollment-test mismatch, as detailed below:

\begin{itemize}


\item \textbf{Cross-Channel test}: Conducted on the AIShell-1 dataset,
where the enrollment and test data are recorded at the same time but with different recording devices.
It tests the performance with physical channel mismatch.

\item \textbf{Time-Variance test}: Conducted on the CSLT-Chronos dataset,
where the test data are recorded several times, and weeks or months later than the enrollment data.
The data are recorded with the same device and in the same acoustic environment.
It tests the performance with long-term speaking behavior mismatch.

\item \textbf{Near-Far test}: Conducted on the HI-MIA dataset,
where the enrollment and test data are recorded at the same time, by devices of the same brand, but
with different recording distances. It tests the performance with recording condition mismatch.

\end{itemize}

\subsection{Data}

Four datasets were used in our experiments: VoxCeleb~\cite{chung2018voxceleb2,nagrani2017voxceleb},
AIShell-1~\cite{aishell_2017}, CSLT-Chronos~\cite{wang2010creation}, and HI-MIA~\cite{himia}.
VoxCeleb was used to train the speaker embedding model, which is the x-vector model in our experiment.
The other three datasets were used for performance evaluation under different enrollment-test mismatch scenarios.
More information about these datasets is presented below.

\emph{VoxCeleb\footnote{http://www.robots.ox.ac.uk/\textasciitilde vgg/data/voxceleb/}}: This is a large-scale audiovisual speaker dataset collected by the University of Oxford~\cite{chung2018voxceleb2}.
The entire database consists of \emph{VoxCeleb1} and \emph{VoxCeleb2}.
All the speech signals were collected from open-source media channels and therefore involved rich variations in channel, style, and ambient noise.
This dataset was used to train the x-vector embedding model.
The entire dataset contains $2,000$+ hours of speech signals from $7,000$+ speakers.
Data augmentation was applied to improve robustness, with the MUSAN corpus~\cite{musan2015} used to generate noisy utterances,
and the room impulse responses (RIRS) corpus~\cite{ko2017study} was used to generate reverberant utterances.

\emph{AIShell-1}\footnote{http://openslr.org/33/}: This is an open-source multi-channel Chinese Mandarin speech dataset published by AISHELL~\cite{aishell_2017}.
All the speech utterances are recorded in parallel via three categories of devices,
including high fidelity Microphones (Mic), Android phones (AND) and Apple iPhones (iOS).
This dataset is used for the \textbf{Cross-Channel test} in our experiment.
The entire dataset consists of two parts: \emph{AIShell-1.Train}, which covers $3$ devices and involves $360,897$ utterances from $340$ speakers,
was used to implement both the ad-hoc normalization/adaptation methods and our proposed SD/LT scoring methods.
\emph{AIShell-1.Eval}, which also covers $3$ devices and involves $64,495$ utterances from $60$ speakers,
was used for performance evaluation on different enrollment-test conditions.
For each cross-channel test, it consists of $21$k target trials and $1,250$k imposter trials.

\emph{CSLT-Chronos\footnote{http://data.cslt.org/}}: This is a time-varying speaker recognition dataset collected by Tsinghua University~\cite{wang2010creation,wang2016improving}. For commercial usage, only the x-vectors are published for research usage.
There are $8$ recording sessions selected to evaluate the time-variance effect,
and all these sessions have the same acoustic environment (speakers, prompt texts and recording devices).
The first $4$ sessions are recorded in an interval of approximately one week,
and the following $4$ sessions are recorded in an interval of approximately one month.
This dataset is used for the \textbf{Time-Variance test} in our experiment.
In each session, it involves $57$ speakers, and each speaker has $150$ utterances.

For each speaker, the first $20$ utterances in each recording session are used as the enrollment set
as well as the development set to train the scoring model, and the rest $130$ utterances are used for evaluation.
Note that this is not a typical setting as the same speakers are used for both model training and evaluation (enrollment part).
We had to choose such a configure as the number of speakers in CSLT-Chronos is limited.
During the test, we will enroll the speakers with data of one session and verify them with data of multiple sessions.
As more sessions involved in the test (from $1$ to $8$), the number of trials increases from $421$k to $3,377$k.

\emph{HI-MIA}\footnote{http://openslr.org/85/}: This is an open-source far-field text-dependent speaker recognition dataset published by AISHELL~\cite{himia}.
This dataset is used for AISHELL Speaker Verification Challenge 2019\footnote{http://challenge.aishelltech.com/}.
All the speech utterances are recorded with the wake-up words `Hi-MIA' from $1$m, $3$m and $5$m fields in parallel.
This dataset is used for the \textbf{Near-Far test} in our experiment.
Specially, the recordings in the normal speaking speed from the clean recording environment are selected in our experiments.
Finally, we build two subsets: \emph{HI-MIA.Train}, which covers $3$ fields and involves $15,186$ utterances from $254$ speakers,
was used to implement different scoring methods.
\emph{HI-MIA.Eval}, which also covers $3$ fields and involves $5,048$ utterances from $86$ speakers,
was used for performance evaluation.
For each near-far test, there are about $1.2$k target trials and $100$k imposter trials.

\subsection{Embedding models}

We built the state-of-the-art x-vector embedding model based on the TDNN architecture~\cite{snyder2018xvector}.
This x-vector model was created using the Kaldi toolkit~\cite{povey2011kaldi}, following the VoxCeleb recipe.
The acoustic features are $40$-dimensional Fbanks. The main architecture contains three components.
The first component is the feature-learning component, which involves $5$ time-delay (TD) layers to learn frame-level speaker features.
The splicing parameters for these $5$ TD layers are:
\{$t$-$2$, $t$-$1$, $t$, $t$+$1$, $t$+$2$\}, \{$t$-$2$, $t$, $t$+$2$\}, \{$t$-$3$, $t$, $t$+$3$\}, \{$t$\}, \{$t$\}.
The second component is the statistical pooling component,
which computes the mean and standard deviation of the frame-level features from a speech segment.
The final one is the speaker-classification component, which discriminates different speakers.
This component has $2$ full-connection (FC) layers and the size of its output is $7,185$,
corresponding to the number of speakers in the training set.
Once trained, the $512$-dimensional activations of the penultimate FC layer are read out as an x-vector.

\subsection{Statistical analysis}

\begin{figure}[htb]
  \centering
  \includegraphics[width=0.9\linewidth]{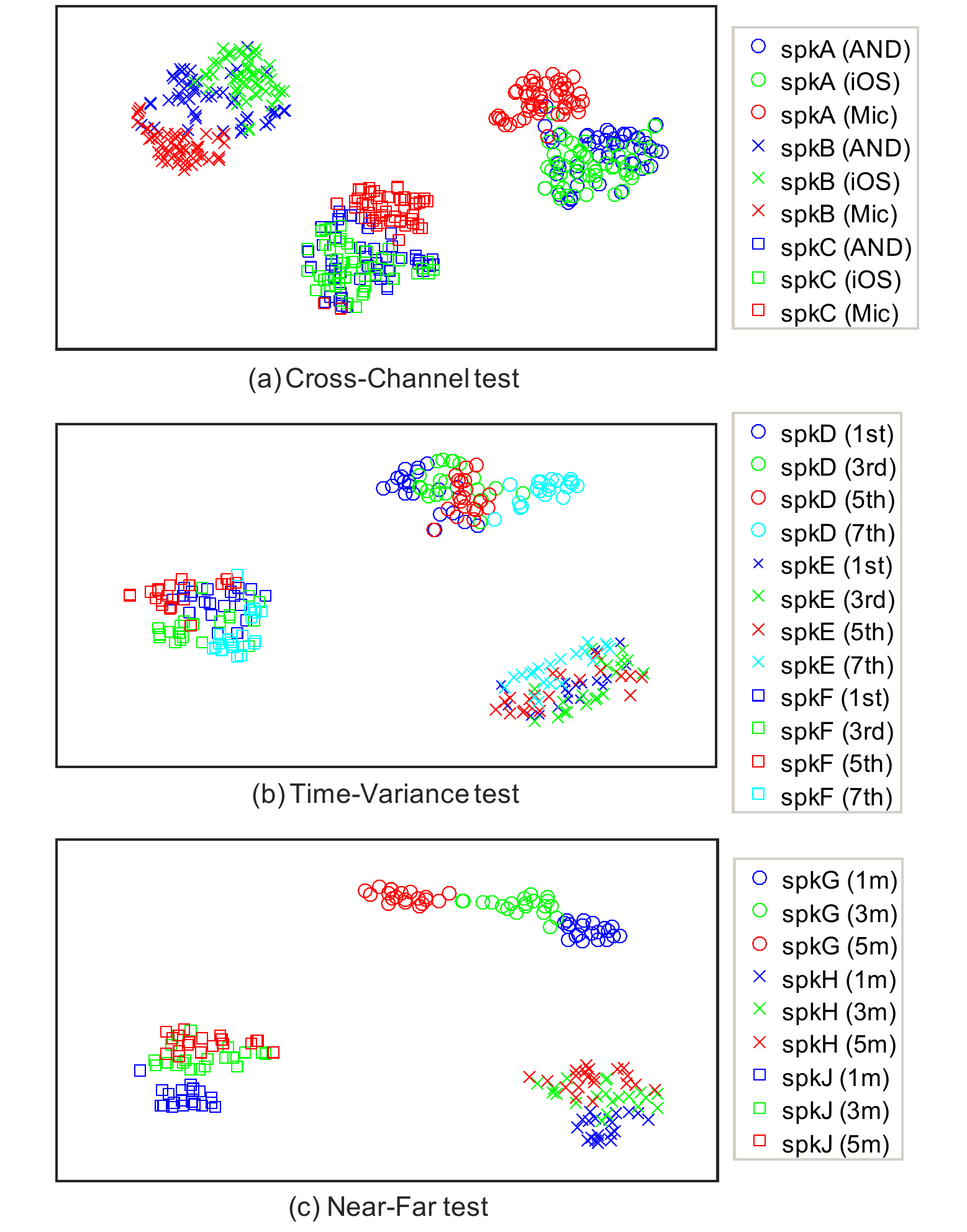}
  \caption{The distributions of x-vectors in three test scenarios. Each shape represents one speaker and each color represents a condition, which is
  (a) device in the Cross-Channel test, (b) recording session in the Time-Variance test, and (c) recording distance in the Near-Far test.}
  \label{fig:tsne}
\end{figure}

In this section, we conduct a quantitative analysis for the statistics incoherence problem associated with the three enrollment-test mismatch tests.
To gain an intuitive understanding for the impact of different mismatch scenarios, we firstly visualize the x-vectors of
three randomly chosen speakers for each scenario, using the t-SNE algorithm~\cite{saaten2008}. As shown in Fig.~\ref{fig:tsne},
in all the three mismatch scenarios, the distribution of x-vectors of each single speaker changes clearly from one condition to another.

We further will evaluate some statistical quantities for data in different conditions and analyze the differences on the quantities from one condition
to another. The tested quantities are explained as follows:

\begin{itemize}

\item Between-speaker and within-speaker variances. Since the covariance matrices (either between-speaker or within-speaker)
      of different conditions may be in different directions, we test the variance in the principle directions of each condition.
      Specifically, we select one condition and estimate an LDA transformation, and then project x-vectors of the test speech in
      all the conditions (including the enrollment condition itself) to the LDA space. Finally, we plot
      the between-speaker and within-speaker variances on dimensions that correspond to the $50$ leading directions of the
      LDA transformation.

\item Angle metric and length metric to measure the global shift.
     The global mean vectors of x-vectors in the enrollment condition $\pmb{x}_e$ and the test condition $\pmb{x}_t$ are firstly computed,
     and then the cosine distance and Euclidean distance between them are computed as the angle metric and length metric
     to measure the global shift. The angle metric is formulated by:
    \begin{equation}
     \label{eq:angle}
     \theta = (1 - \frac{\pmb{x}_e^{T}  \pmb{x}_t}{||\pmb{x}_e|| ~ ||\pmb{x}_t||}) * 10^{3},
    \end{equation}
    and the length metric is formulated by:
    \begin{equation}
     \label{eq:length}
     \ell = ||\pmb{x}_e - \pmb{x}_t||^2 * 10^{2}.
    \end{equation}

\end{itemize}

\subsubsection{Cross-Channel test}

We use \emph{AIShell-1.Train} to compute the statistics. The dataset involves $340$ speakers recorded with $3$ recording devices at the same time.

\begin{figure}[!htb]
    \centering
    \includegraphics[width=1\linewidth]{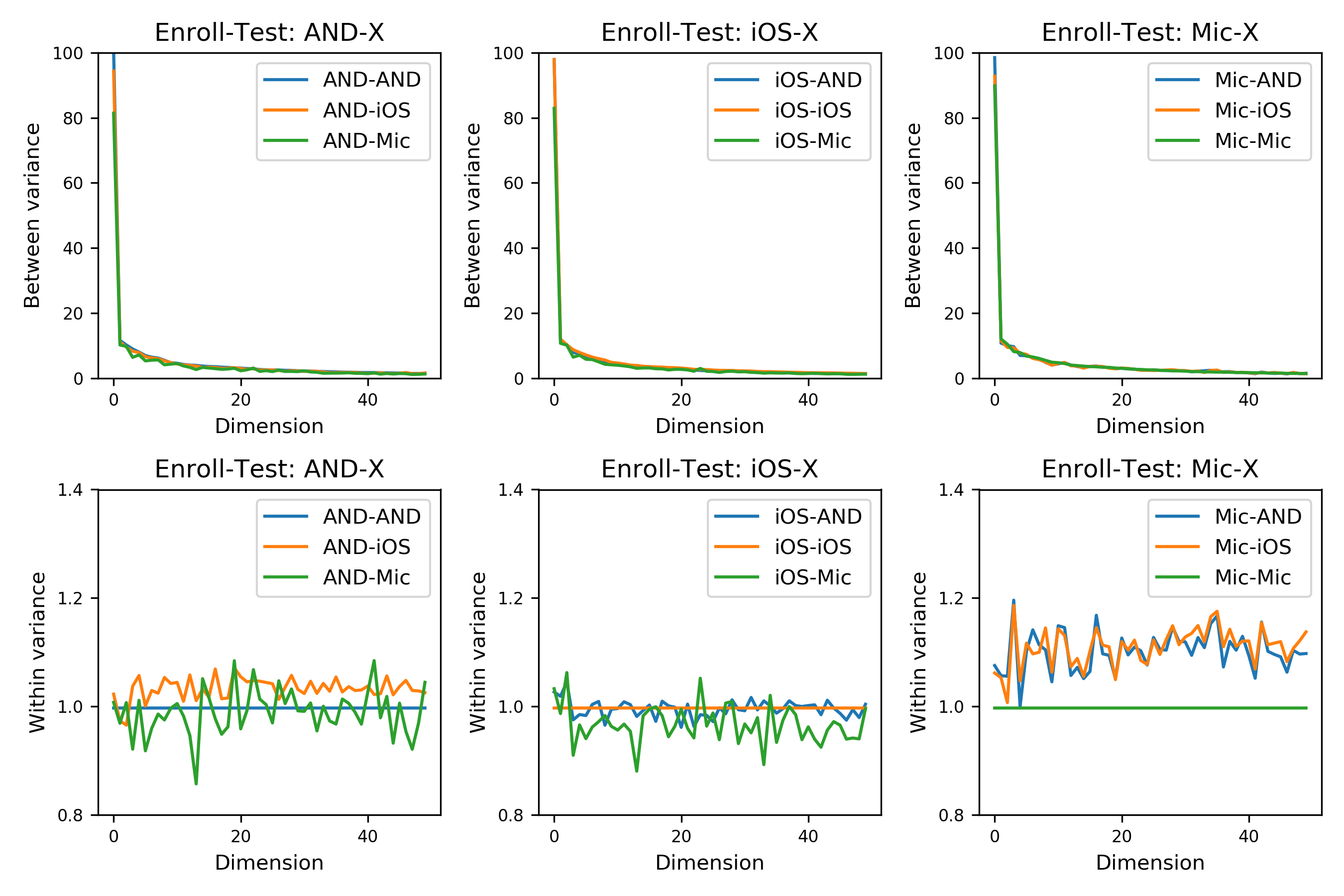}
    \caption{Between-speaker and within-speaker variances (the first $50$ dimensions) in the Cross-Channel test.}
    \label{fig:channel}
\end{figure}

The between-speaker and within-speaker variances of the test data are reported in Figure~\ref{fig:channel}, where
each enrollment condition is presented separately.
The observations are: (1) The pattern of the between-speaker variances are not
significantly different for different enrollment conditions, and for each enrollment condition, the between-speaker variances for test in
different conditions are also similar. (2)
For each of the enrollment condition, the pattern of within-speaker variances is different for test in different conditions,
but the difference is rather marginal. The within-class variances under enrollment-test mismatched conditions are in the range of $0.8$ to $1.2$,
which is close to $1.0$ that is the optimal value under enrollment-test matched conditions (e.g., Mic-Mic)\footnote{LDA plays a role of distribution normalization,
by which the within-speaker covariance is normalized to an identity matrix $\mathbf{I}$.}.

The angle metric and length metric that measure the global shift are reported
in Table~\ref{tab:channel}, where we report the results with all pairs of
mismatched conditions. Note that for global shift, the mismatch is symmetric for
enrollment and test.
It can be observed that the global shift is evident in all the mismatched conditions, and
the shift between AND and iOS is much smaller than that between Mic and AND/iOS.

All the results coincide the observations in the qualitative analysis,
and suggest that in the Cross-Channel test, the enrollment-test mismatch
is mainly reflected in a global shift. Therefore, it can be compensated for by a simple
shifting back, as formulated by the first special case of the SD/LT approach.

\begin{table}[!htb]
  \caption{Angle metric and length metric in the Cross-Channel test.}
  \label{tab:channel}
  \centering
  \scalebox{1.3}{
  \begin{tabular}{lcc}
  \cmidrule(r){1-1} \cmidrule(r){2-3}
      Enroll-Test   &  Angle ($\theta$) &  Length ($\ell$)  \\
  \cmidrule(r){1-1} \cmidrule(r){2-3}
      AND-iOS       &  0.940            &  4.673            \\
      iOS-Mic       &  5.199            &  10.789           \\
      Mic-AND       &  5.645            &  10.867           \\
  \cmidrule(r){1-1} \cmidrule(r){2-3}
  \end{tabular}}
\end{table}

\subsubsection{Time-Variance test}

We use the \emph{CSLT-Chronos} dataset to measure the statistics incoherence in the Time-Variance test.
The dataset involves $57$ speakers recorded with $8$ recording sessions in the same acoustic environment.
The data of the first session is used for enrollment, and all the $8$ sessions are used for test.
In order to study the incremental change on the statistics, we perform a serial test where
the test sessions are gradually added to the test data. For example, the test \textbf{1st-[1st:5th]} means the enrollment
is based on the first session, and the test is based on the data from the first to the fifth sessions.
Note that the test \textbf{1st-[1st:1st]} is the condition-matched test.

\begin{figure}[!htb]
    \centering
    \includegraphics[width=\linewidth]{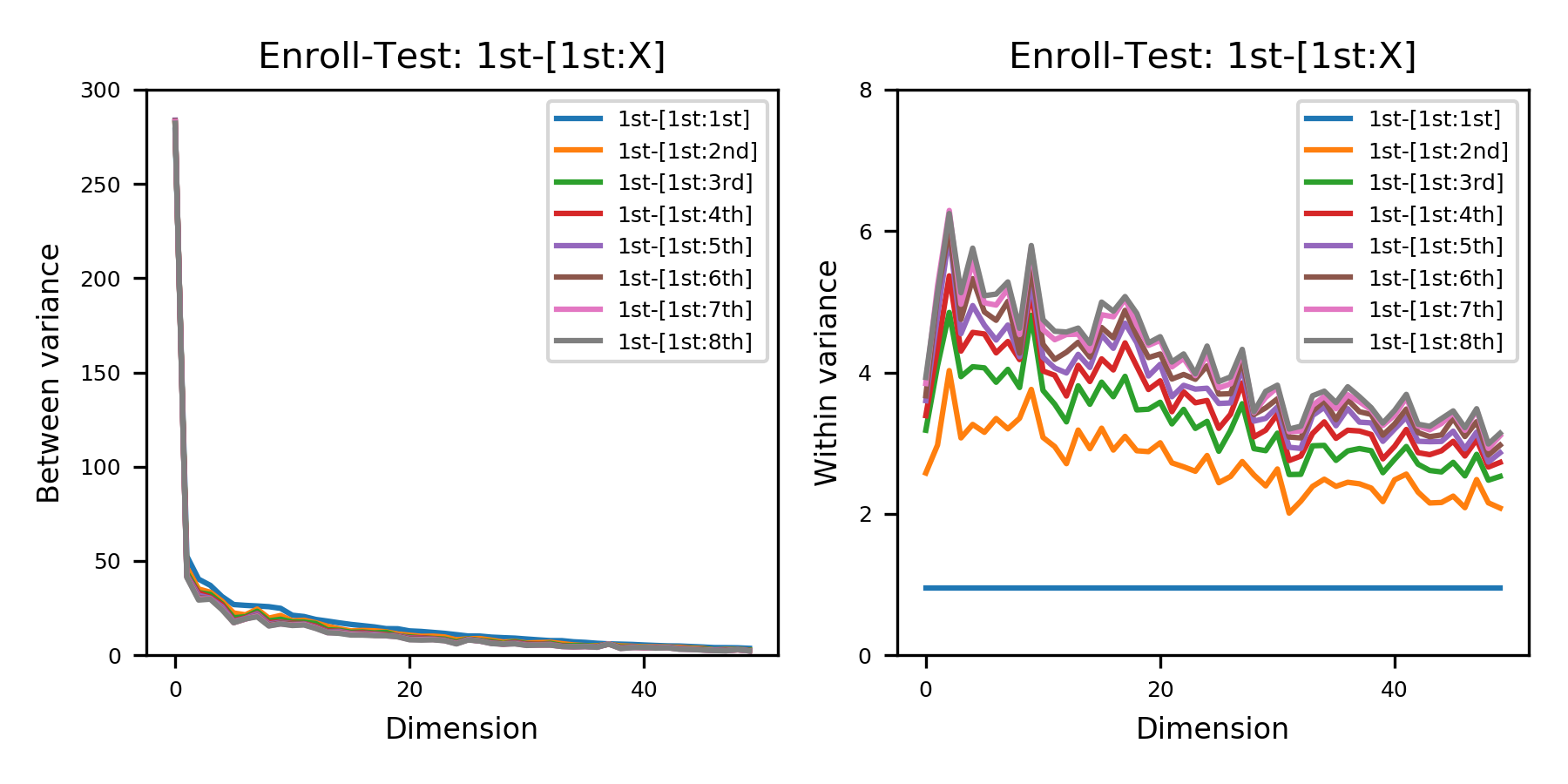}
    \caption{Between-speaker and within-speaker variances (the first $50$ dimensions) in the Time-Variance test.}
    \label{fig:time}
\end{figure}

The between-speaker and within-speaker variances of the test data
are reported in Figure~\ref{fig:time}, where the enrollment is based on the first session in all the test cases.
It can be observed that with more sessions involved in the test data,
the pattern of the between-speaker variances does not change much,
but the within-speaker variances are gradually increased and finally converge to stable values.
The angle metric and length metric for the condition-mismatched tests are reported in Table~\ref{tab:time}.
Compared to the values in the Cross-Channel test shown in Table~\ref{tab:channel},
one can find that the global shift is much smaller in the Time-Variance test.

\begin{table}[htb!]
  \caption{Angle metric and length metric for global shift in the Time-Variance test.}
  \label{tab:time}
  \centering
  \scalebox{1.3}{
  \begin{tabular}{lcc}
  \cmidrule(r){1-1} \cmidrule(r){2-3}
      Enroll-Test   &  Angle ($\theta$) &  Length ($\ell$) \\
  \cmidrule(r){1-1} \cmidrule(r){2-3}
       1st-[1st:2nd]      &  0.085            &  1.615   \\
       1st-[1st:3rd]      &  0.159            &  2.326   \\
       1st-[1st:4th]      &  0.234            &  2.654   \\
       1st-[1st:5th]      &  0.269            &  2.991   \\
       1st-[1st:6th]      &  0.301            &  3.224   \\
       1st-[1st:7th]      &  0.346            &  3.439   \\
       1st-[1st:8th]      &  0.393            &  3.590   \\
  \cmidrule(r){1-1} \cmidrule(r){2-3}
  \end{tabular}}
\end{table}

All these observations are consistent with the results obtained from the qualitative analysis presented in the previous section.
From these observations, we can conclude that in the Time-Variance test, the enrollment-test mismatch mainly impacts the within-speaker
variance. Therefore, a simple within-speaker variance adaptation may be a good approach to alleviate the mismatch, as
assumed by the second special case of the SD/LT approach.

\subsubsection{Near-Far test}

The \emph{HI-MIA.Train} dataset is used to compute the statistics in the Near-Far test.
It involves $254$ speakers recorded with $3$ recording distances at the same time.

\begin{figure}[!htb]
    \centering
    \includegraphics[width=\linewidth]{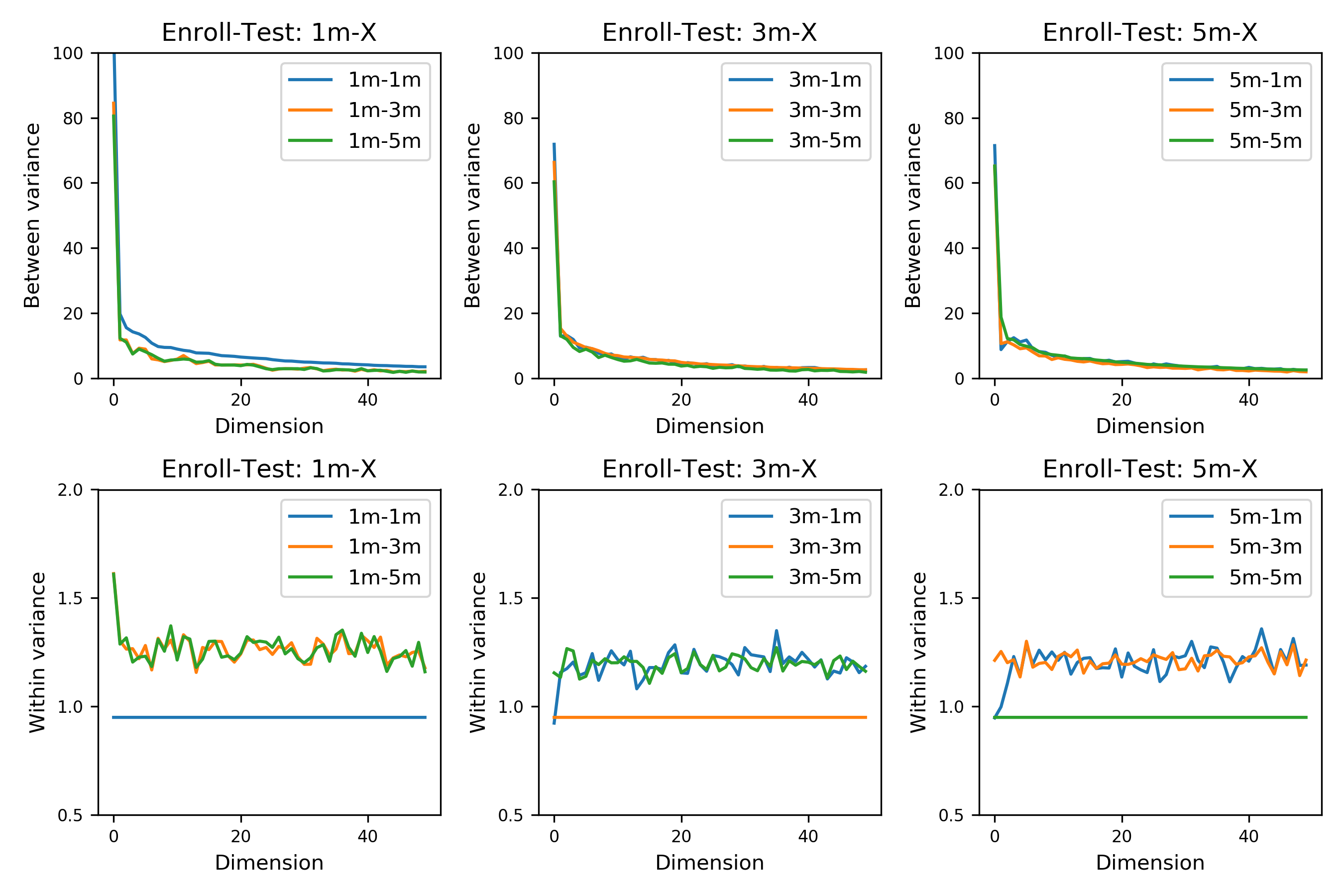}
    \caption{Between-speaker and within-speaker variances (the first 50 dimensions) in the Near-Far test.}
    \label{fig:nf}
\end{figure}

Similar to in the Cross-Channel test, we examine the between-speaker and within-speaker variances
of the test data and present them independently on three enrollment conditions.
The results are shown in Figure~\ref{fig:nf}.
Firstly, it can be observed that the patterns of the between-speaker variances with the three
enrollment conditions are different: enrolling with the near-filed data ($1$m) shows a relatively
large between-speaker variance. Moreover, in each enrollment condition, the near-field test data
shows a higher between-speaker variance. These observations demonstrated that: (1) the near-field data
is more speaker discriminative on either the enrollment or test condition; (2) the mismatch between
the near-field and far-field data is reflected in the between-speaker variance.

Paying attention to the within-speaker variance, it shows that in all the enrollment conditions, the
condition-matched test data (e.g., $3$m-$3$m) show the smallest and theoretically correct
within-speaker variance; all the condition-mismatched test data exhibit larger within-speaker variances.
It suggests that the within-speaker variances in different conditions are different.

The angle metric and length metric are reported in Table~\ref{tab:nf}.
It can be observed that the global shifts between two different conditions are significant, when
compared to the results in Cross-Channel test (Table~\ref{tab:channel}) and Time-Variance test (Table~\ref{tab:time}).
Moreover, the global shift between the two far-field conditions is much smaller than that between the near-field condition
and the two far-field conditions.

As a summary, in the Near-Far test, the statistics incoherence is highly complex. It involves
mismatch in both the between-speaker and within-speaker variances, as well as a clear global shift.
Therefore, it cannot be well addressed by
the two special cases of the SD/LT approach; the general form of SD/LT should be employed.

\begin{table}[htb!]
  \caption{Angle metric and length metric in the Near-Far test.}
  \label{tab:nf}
  \centering
  \scalebox{1.3}{
  \begin{tabular}{lcc}
  \cmidrule(r){1-1} \cmidrule(r){2-3}
      Enroll-Test   &  Angle ($\theta$) &  Length ($\ell$) \\
  \cmidrule(r){1-1} \cmidrule(r){2-3}
        1m-3m       &  4.959            &  10.745          \\
        1m-5m       &  6.509            &  11.890          \\
        3m-5m       &  1.703            &  4.319           \\
  \cmidrule(r){1-1} \cmidrule(r){2-3}
  \end{tabular}}
\end{table}

\subsection{Scoring methods in comparison}

We verify the SD/LT method on the speaker recognition task, by comparing different forms of SD/LT (two special cases and one general form)
and several baseline methods.
To make the comparison fair, we categorize these methods by the data it requires: the no-data methods
do not require any data from the test condition; the unparallel-data methods require data from the test condition but the data are
independent from the enrollment data; the parallel-data methods require data from the test condition and the speakers are shared with
the enrollment data.

All the methods are based on the PLDA scoring, with the PLDA models~\cite{Ioffe06} implemented using the Kaldi toolkit~\cite{povey2011kaldi}.
The SD/LT source code has been published online\footnote{https://gitlab.com/csltstu/enroll-test-mismatch}.

\subsubsection{Non-data methods}

\begin{itemize}
\item \textbf{Baseline}: The basic PLDA scoring approach where the statistical models are trained with development set in the enrollment condition,
and the model is employed to perform the test by ignoring the enrollment-test mismatch.
\end{itemize}

\subsubsection{Nonparallel-data methods}

\begin{itemize}
\item \textbf{IDVC}: Inter-dataset variability compensation (IDVC). It is a popular unsupervised domain normalization technique~\cite{aronowitz2014inter},
  and has shown great advantage especially in cross-channel scenarios~\cite{kanagasundaram2015improving}.
  In our experiments, IDVC is implemented using the nuisance attribute projection (NAP) method~\cite{solomonoff2007nuisance}.
  We combine the development data from both enrollment and test conditions to estimate the IDVC projection.
  For each of the two conditions, the data is split to $n$ subsets and the center vectors of all the subsets are computed.
  NAP is then performed with the 2$n$ center vectors ($n$ from enroll and $n$ from test).
  We experimented various configurations and found the optimal $n$ is 2, 20, 5 for the cross-channel test, time-variance test and near-far test, respectively.
  The optimal $n$ is used in the evaluation.

  \item \textbf{GSC}: Global shift compensation (GSC). It corresponds to the special case 1 of the SD/LT approach, as mentioned in Section~\ref{sec:gsc},
  and assumes that the enrollment-test mismatch can be largely compensated for by a global shift, as shown in Eq.(\ref{eq:nl-gsc}). Note that the global shift can be
  easily computed by the nonparallel development data.
  Due to the previous qualitative and quantitative analysis,  we expect that this method is effective in the Cross-Channel test.

  \item \textbf{WVA}: Within variance adaptation (WVA). It corresponds to the special case 2 of the SD/LT approach, as mentioned in Section~\ref{sec:wva}.
  It assumes that the enrollment-test mismatch is mainly due to the difference in the within-speaker variance, and so can be
  compensated for by using the correct within-speaker variance as Eq.~(\ref{eq:nl-wva}). Note that the within-speaker variances in the enrollment and test
  conditions can be obtained using the nonparallel development data.
  Due to the previous qualitative and quantitative analysis, we expect this method is effective in the Time-Variance test.
\end{itemize}

\subsubsection{Parallel-data methods}

\begin{itemize}

  \item \textbf{SD/LT}: Statistics decomposition with linear transformation (SD/LT). It corresponds to the general form as mentioned in Section~\ref{sec:gf}.
  This model is a linear transformation $\pmb{x} = \mathbf{M} \hat{\pmb{x}} + \pmb{b}$
  that maps the test sample $\hat{\pmb{x}}$ to the enrollment condition so that $\hat{\pmb{x}}$ can be represented by the
  statistical model of the enrollment condition. Importantly, the transformation is only used to compute
  the likelihood term $p_k(\pmb{x})$; and the normalization term $p(\pmb{x})$ is still computed using
  the statistical model of the test condition, based on the original test data $\hat{\pmb{x}}$.
  We implemented it based on the MLE criterion with the training objective shown in Eq.~(\ref{eq:mle-loss}).
  In our experiment, the Adam optimizer~\cite{kingma2014adam} was used to optimize the parameters $\{\mathbf{M}, \pmb{b}\}$.

  \item \textbf{MCT}: Multi-condition training (MCT). A widely used baseline approach to deal with
  enrollment-test condition mismatch.
  In this method, data from both the enrollment and test conditions are pooled together,
  and then are used to train the PLDA model. This model can be regarded as an interpolation
  of the statistical models for the enrollment and test conditions. Although not
  an optimal solution, MCT often delivers good performance in practice. Note that it requires
  parallel data to discover the speaker subspace shared by different conditions.

  \item \textbf{CAT}: Condition-adaptation training (CAT). Another simple baseline that
  transforms speaker vectors from the test condition to the enrollment condition and
  then performs the scoring there. The transformation is linear and is the same as in SD/LT,
  defined by $\pmb{x} = \mathbf{M} \hat{\pmb{x}} + \pmb{b}$, and the
  parameters $\{\mathbf{M}, \pmb{b}\}$ are estimated using the same ML training
  as the SD/LT approach. We highlight that the main difference between
  SD/LT and CAT is that for SD/LT, the normalization term $p(\pmb{x})$ of the LLR score is computed using the model of the
  test condition; while for CAT, it is computed using the model of the enrollment condition.
  This difference is not trivial: the normalization term computed in the test condition is optimal if the
  statistical model for the test condition is accurate, so SD/LT will be MBR optimal under that condition, but
  CAT is not.

\end{itemize}

\subsection{Speaker recognition results}

In this section, experimental results in terms of equal error rate (EER) are
presented for the three speaker recognition tests: the Cross-Channel test,
the Time-Variance test and the Near-Far test. We will use `Base', `NPD', `PD' to
represent the Baseline method, the nonparallel-data methods (including GSC and WVA),
and the parallel-data methods (including SD/LT, MCT and CAT), respectively.
We also report the performance with AS-norm applied upon SD/LT scores, to show the
potential complementarity of the two approaches. For comparison, the results of the
baseline system with AS-norm are also reported.

\subsubsection{Cross-Channel test}

In this test, the statistical models of all the scoring methods are trained with the \emph{AIShell-1.Dev} dataset,
and the results on the \emph{AIShell-1.Eval} dataset are reported in Table~\ref{tab:channel-test}. The observations
are as follows.

\begin{table*}[htb!]
  \caption{EER(\%) on the Cross-Channel test.}
  \label{tab:channel-test}
  \centering
  \scalebox{1.08}{
  \begin{tabular}{lccccccccc}
     \cmidrule(r){1-10}
       \multirow{2}{*}{Enroll-Test} & \multirow{2}{*}{Base} &  \multirow{2}{*}{Base + AS-norm}    & \multicolumn{3}{c}{NPD} & \multicolumn{3}{c}{PD} \\
                      \cmidrule(r){4-6} \cmidrule(r){7-10}
                   & &             &  IDVC    &  GSC     &  WVA     &  MCT     &  CAT      &  SD/LT & SD/LT + AS-norm      \\
      \cmidrule(r){1-1} \cmidrule(r){2-2} \cmidrule(r){3-3} \cmidrule(r){4-6} \cmidrule(r){7-10}
         AND-AND   &   0.797     & 0.684 &  -        &  -       &  -       &  -       &  -        &  -        & - \\
         AND-Mic   &   2.146     & 1.316 &  1.768    &  1.764   &  2.165   &  1.151   &  1.245    &   0.981 & \textbf{0.943}     \\
         AND-iOS   &   1.425     & 0.741 &  1.354    &  1.382   &  1.401   &  1.161   &  1.312    &  \textbf{0.623} & 0.760    \\
      \cmidrule(r){1-1} \cmidrule(r){2-2} \cmidrule(r){3-3} \cmidrule(r){4-6} \cmidrule(r){7-10}
         Mic-AND   &   2.175     & 1.382 &  1.665    &  1.665   &  2.033   &  1.161   &  1.189    &  \textbf{0.712} & 1.128    \\
         Mic-Mic   &   0.778     & 0.849 &  -        &  -       &  -       &  -       &  -        &  -              & - \\
         Mic-iOS   &   2.251     & 1.231 &  1.920    &  1.892   &  2.081   &  1.293   &  1.481    &  \textbf{0.812} & 1.038    \\
      \cmidrule(r){1-1} \cmidrule(r){2-2} \cmidrule(r){3-3} \cmidrule(r){4-6} \cmidrule(r){7-10}
         iOS-AND   &   1.599     & 0.972 &  1.382    &  1.430   &  1.590   &  1.156   &  1.184    &  \textbf{0.755} & 0.854    \\
         iOS-Mic   &   2.216     & 1.330 &  1.726    &  1.759   &  2.231   &  1.137   &  1.231    &  1.052          & \textbf{0.948}    \\
         iOS-iOS   &   0.920     & 0.684 &  -        &  -       &  -       &  -       &  -        &  -        & - \\
    \cmidrule(r){1-10}
  \end{tabular}}
\end{table*}

\begin{itemize}

\item The performance with enrollment-test matched conditions is much
higher than those with mismatched conditions, which indicates that the enrollment-test mismatch is indeed a
serious problem for speaker recognition. Moreover, the performance loss with different mismatched condition
pairs is different. For instance, The pair AND-iOS obtains better performance than the pair AND-Mic.
This is consistent to the observations in the qualitative and quantitative analysis,
in particular the results in Figure~\ref{fig:channel} and Table~\ref{tab:channel}.

\item For the nonparallel-data methods, IDVC and GSC consistently outperform the baseline on all these mismatched conditions,
while WVA does not show much advantage.
This observation supports our argument that the main effect of the cross-channel mismatch
is a global shift, and even a simple GSC can compensate for the mismatch to a large extent.
Besides, it can be found that IDVC is superior to GSC.
This indicates dataset-invariant projection by IDVC
works better than simple mean subtraction by GSC.

\item The three parallel-data methods deliver much better performance than the
two nonparallel-data methods.
That is not surprising as the parallel-data methods utilize extra information of cross-channel speakers.
Moreover, the proposed SD/LT approach consistently outperforms MCT and CAT, demonstrating that SD/LT is
a more effective approach in dealing with enrollment-test mismatch. The comparison between SD/LT and CAT
is especially interesting, as the two methods look very similar and only differ in the normalization term. The clear
advantage of SD/LT demonstrated the importance of a principle solution based on a solid theory.

\item The AS-norm consistently improves the baseline results, confirming its effectiveness. However, when applied together with SD/LT,
no consistent improvement is observed. A possible reason is that the SD/LT performance has been sufficiently optimal, leaving
little space for the AS-norm to make contribution.

\end{itemize}

\subsubsection{Time-variance test}

The test is conducted with the \emph{CSLT-Chronos} dataset, which consists of 57 speakers and each speaker has 150 utterances.
The first 20 utterances of each speaker are selected as the enrollment set, as well as the development set to train the scoring methods.
The rest 130 utterances are used for evaluation.
The EER results are reported in Table~\ref{tab:time-test}.
The observations are as follows.

\begin{table*}[htb!]
  \caption{EER(\%) on the Time-Variance test.}
  \label{tab:time-test}
  \centering
  \scalebox{1.08}{
  \begin{tabular}{lccccccccc}
     \cmidrule(r){1-10}
       \multirow{2}{*}{Enroll-Test} & \multirow{2}{*}{Base} &  \multirow{2}{*}{Base + AS-norm}    & \multicolumn{3}{c}{NPD} & \multicolumn{3}{c}{PD} \\
                      \cmidrule(r){4-6} \cmidrule(r){7-10}
                   & &             &  IDVC    &  GSC     &  WVA     &  MCT     &  CAT      &  SD/LT & SD/LT + AS-norm      \\
      \cmidrule(r){1-1} \cmidrule(r){2-2} \cmidrule(r){3-3} \cmidrule(r){4-6} \cmidrule(r){7-10}
         1st-[1st:1st]  &  4.799   &  5.016 &  -        &  -       &  -       &  -       &  -        &  -             & - \\
         1st-[1st:2nd]  &  6.400   &  6.306 &  6.258    &  6.346   &  5.934   &  5.258   &  5.549    &  \textbf{4.339}&  5.218   \\
         1st-[1st:3rd]  &  6.863   &  6.282 &  6.669    &  6.777   &  6.156   &  4.976   &  5.233    &  \textbf{4.062}&  5.111   \\
         1st-[1st:4th]  &  6.884   &  6.256 &  6.850    &  6.810   &  6.084   &  4.619   &  4.882    &  \textbf{3.710}&  4.612   \\
         1st-[1st:5th]  &  7.108   &  6.309 &  6.846    &  7.022   &  6.230   &  4.348   &  4.804    &  \textbf{3.678}&  4.413   \\
         1st-[1st:6th]  &  7.856   &  6.832 &  7.595    &  7.768   &  6.938   &  4.348   &  4.861    &  \textbf{3.661}&  4.298   \\
         1st-[1st:7th]  &  7.906   &  7.015 &  7.618    &  7.825   &  7.005   &  4.262   &  4.903    &  \textbf{3.749}&  4.473   \\
         1st-[1st:8th]  &  7.881   &  7.034 &  7.737    &  7.815   &  6.993   &  4.300   &  5.041    &  \textbf{3.937}&  4.499   \\
    \cmidrule(r){1-10}
  \end{tabular}}
\end{table*}

\begin{itemize}

\item With more sessions involved in the test data, the performance of the baseline system
is gradually reduced, and finally tends to converge to a stable value. This is consistent with our
qualitative and quantitative analysis that with more sessions involved, the within-speaker variance
will be gradually increased, leading to more severe enrollment-test mismatch.

\item Looking at the performance of the three nonparallel-data methods, it can be found that
compared with the baseline, WVA obtains obvious performance improvement in all the test conditions,
while IDVC/GSC does not show much help. This supports our argument that the main effect of the time-variance
mismatch is on the within-speaker variance.

\item Again, the three parallel-data methods attain  much better performance than the two nonparallel-data methods,
and SD/LT attains the best performance. Interestingly, when the test data is sufficiently large, the SD/LT performance is
even better than the condition-matched baseline, i.e., the test \textbf{1st-[1st:1st]}. We attribute this good performance
to the more reliable estimation for the within-speaker and between-speaker variances
of the test condition. As a comparison, the CAT performance is never better than the condition-matched baseline,
as it does not fully use the statistical model of the test condition.

\item The AS-norm slightly improves the baseline results for scenarios with enrollment-test mismatch, but
impairs the performance for scenarios without such mismatch. The performance loss is also observed in the results applied upon SD/LT.
This may be attributed to the limited cohort speakers in the test set (57 speakers in total), which prevents reliable estimation for
the score distribution. This imperfect estimation may worsen the performance if the score normalization is not very necessary, e.g.,
if no enrollment-test mismatch exists, or if the mismatch has been compensated for by SD/LT.

\end{itemize}

\subsubsection{Near-Far field test}

In this test, the \emph{HI-MIA.Dev} dataset is used to train the scoring methods, and
the results on the \emph{HI-MIA.Eval} dataset are reported in Table~\ref{tab:nf-test}.
The observations are as follows.

\begin{table*}[htb!]
  \caption{EER(\%) on the Near-Far test.}
  \label{tab:nf-test}
  \centering
  \scalebox{1.08}{
  \begin{tabular}{lccccccccc}
     \cmidrule(r){1-10}
       \multirow{2}{*}{Enroll-Test} & \multirow{2}{*}{Base} &  \multirow{2}{*}{Base + AS-norm}    & \multicolumn{3}{c}{NPD} & \multicolumn{3}{c}{PD} \\
                      \cmidrule(r){4-6} \cmidrule(r){7-10}
                   & &             &  IDVC    &  GSC     &  WVA     &  MCT     &  CAT      &  SD/LT & SD/LT + AS-norm      \\
      \cmidrule(r){1-1} \cmidrule(r){2-2} \cmidrule(r){3-3} \cmidrule(r){4-6} \cmidrule(r){7-10}
         1m-1m  &  0.620     & 0.465    &  -       &  -       &  -       &  -        &  -        &  -       &-  \\
         1m-3m  &  3.968     & 2.672    & 3.725   &  3.725   &  3.644   &  \textbf{2.510}        &  3.563    &  2.996  & 2.591  \\
         1m-5m  &  4.866     & 3.244    & 4.623   &  4.623   &  4.623   &   3.082    &  4.623    &  3.731  & \textbf{2.676}  \\
      \cmidrule(r){1-1} \cmidrule(r){2-2} \cmidrule(r){3-3} \cmidrule(r){4-6} \cmidrule(r){7-10}
         3m-1m  &  1.938     &0.930      &  1.628   &  1.705   &  1.705   &  1.318   &  1.628    &  1.085 & \textbf{0.698}    \\
         3m-3m  &  0.891     &0.648      &  -       &  -       &  -       &  -       &  -        &  -       &-  \\
         3m-5m  &  3.244     &2.028      &  3.244   &  3.325   &  3.082   &  2.595   &  3.244    &  2.433 & \textbf{1.703}   \\
      \cmidrule(r){1-1} \cmidrule(r){2-2} \cmidrule(r){3-3} \cmidrule(r){4-6} \cmidrule(r){7-10}
         5m-1m  &  3.566     &1.938      &  3.333   &  3.256   &  3.411   &  2.326   &  2.481    &  2.171 & \textbf{1.395}   \\
         5m-3m  &  2.834     &\textbf{1.538}      &  2.753   &  2.591   &  2.591   &  2.105   &  2.753    &  2.105 & 1.619    \\
         5m-5m  &  1.135     &1.054     &  -       &  -       &  -       &  -        &  -        &  -      & -   \\
    \cmidrule(r){1-10}
  \end{tabular}}
\end{table*}

\begin{itemize}
\item The performance is seriously degraded with the near-far recording mismatch, especially
when enroll with near-field data and test on far-field data.

\item The three nonparallel-data methods, IDVC, GSC and WVA, all offer reasonable performance improvement compared to the baseline.
This is consistent with our quantitative analysis that this recording condition mismatch incurs both
incoherence on within-speaker variances and global shift.

\item Again, the three parallel-data methods offers more performance gains, as in the Cross-Channel test and the
Time-Variance test. A slight difference is that in this test, SD/LT does not always obtain the best performance.
Among the 6 mismatched conditions, MCT outperforms SD/LT in two conditions. This is a little surprising as SD/LT
is MBR optimal in theory. More careful analysis shows that it is probably because
the nature of linear transformation cannot address complex statistics incoherence caused by the near-far recording mismatch.

\item Finally, the AS-norm improves both the baseline and the SD/LT results. This is a little different from the observation
in the cross-channel/time-variance tests where the AS-norm does not provide consistent improvement upon the SD/LT results. This again may be attributed to the suboptimal
SD/LT results due to the linear transformation.

\end{itemize}

\section{Conclusions}
\label{sec:con}

This paper investigated the issue of enrollment-test mismatch in speaker recognition,
and presented a statistics decomposition (SD) approach to solve this problem.
Specifically, we decouple the scoring process to three separated phases according to
the decomposed formulation of the PLDA scoring, and
statistics from different conditions are used in different phases.
This SD approach guarantees that the resultant scores are MBR optimal,
and therefore is a principle solution for the enrollment-test mismatch problem.

As an initial study, we proposed an implementation of the SD approach based on a
linear transformation (SD/LT).
A comprehensive study was conducted and the experimental results demonstrated that:
(1) Different types of mismatch exhibits different statistics incoherence, and so
should be treated differently;
(2) If the statistics incoherence shows a simple form, a simplified version of SD/LT would
be effective. An advantage of these simplified methods is that they only require
nonparallel data, i.e., data without cross-condition speakers;
(3) If parallel data are available, the general form of SD/LT is highly effective and obtains the best
performance among all the competitive methods.

Future work will investigate the conditional transformation based on more complex transformation functions,
e.g., a deep neural net. This may significantly improve the performance in scenarios
with complex condition mismatch,
for instance the near-far recording mismatch in our experiment. Experiments on diverse enrollment-test
scenarios will be also conducted to demonstrate the strength of the proposed approach, e.g., in the cross-lingual scenario.

\bibliographystyle{IEEEtran}
\bibliography{mybib}

\end{document}